**Altynbek Seitenov**
PhD Researcher in Telemedicine Systems and Health Informatics, L.N. Gumilyov Eurasian National University; Senior Lecturer, Astana IT University, Kazakhstan
Altynbek.Seitenov@astanait.edu.kz
Astana IT University, Kazakhstan
**Azhar Bekbussinova**
B.Sc. Candidate in Software Engineering, Department of Computer Engineering
231552@astanait.edu.kz
Astana IT University, Kazakhstan
**Ainur Nurzhanova**
B.Sc. Candidate in Software Engineering, Department of Computer Engineering
231109@astanait.edu.kz
Astana IT University, Kazakhstan
**Yerassyl Bolatkan**
B.Sc. Candidate in Software Engineering, Department of Computer Engineering
230122@astanait.edu.kz
Astana IT University, Kazakhstan

# Paper Title: HOW CAN EXPLAINABLE AI IMPROVE TRUST AND TRANSPARENCY IN MEDICAL DIAGNOSIS SYSTEMS

**Abstract**: The active adoption of AI in human services has provoked the problem of transparency and trust in decisions, since most healthcare AI systems are black-box models. To resolve these concerns, explainable Artificial Intelligence (XAI) has been suggested as a means of making medical AI tools safer, more reliable, and acceptable through human-readable explanations. This research paper describes the role of XAI in physician trust and acceptance of AI-based diagnostics. The knowledge of XAI, AI decision confidence, perceived usefulness, and adoption intentions were some of the main variables evaluated in a structured survey of 30 medical students and an expert interview. The findings indicate that participants unanimously concurred that AI explanations raise the clarity, safety and acceptability of AI recommendations. The knowledge of XAI showed a strong positive relationship with trust ($r = 0.48$, $p = 0.01$) and perceived usefulness ($r = 0.60$, $p = 0.001$). The model has demonstrated a steady reliability level (Cronbach a = 0.702) and accounted 48-52% variance. Although the study has some limitations, including the small size of the sample and self-reporting, it still has empirical evidence of the positive effect of XAI on human-AI collaboration on the necessary condition of successful integration of AI diagnostic tools in healthcare. Further studies need to be conducted in a clinical inquiry of XAI and determine institutional and patient attitudes.

## Introduction

The rapid advancement in the field of artificial intelligence (AI) over the past several years has led to its penetration in various fields, but healthcare is the sector that has been impacted the most [1]. Deep learning models of machine learning algorithms obtain excellent performance when analyzing medical images and identifying diseases and clinical decision support tasks [2], [3]. The primary issue of the existing AI systems has been that they are uninterpretable black boxes that produce outcomes without revealing how they make decisions [4]. The issue of the absence of transparency in the medical decision-making systems poses significant issues due to their direct impact on patient safety and health outcomes [5]. Lack of transparency in AI systems poses a trust issue to clinicians but obstructs the deployment of systems and presents legal and ethical challenges. The need of Explainable Artificial Intelligence (XAI) has emerged and is now necessary due to the effort to enhance the transparency and accountability and reliability of AI diagnostic systems [6].

The Explainable AI field is a combination of several methods that assist users to learn about the process of decisions made by AI systems and assess their output outcome [2], [7]. Medical diagnosis XAI is better than the correct results since it is able to show clinicians how features influence the decisions and it assists them in finding out the possible biases and comparing the model results with the medical standards. The simplicity of XAI allows improved collaboration between humans and AI that will reduce medical errors and achieve improved patient outcomes [6]. Explainable AI systems are required in healthcare organizations since now regulatory bodies require AI systems to prove interpretability and trustworthiness prior to clinical adoption [1], [8].

The medical AI systems need the utmost trustfulness of the users due to their urgency. Medical workers should realize that system recommendations are based on reliable trends, rather than arbitrary associations or discriminating data [2], [6]. The patients must be assured that AI-based medical decisions ensure their safety and do not interfere with their rights and ethical principles [1]. The existing AI frameworks that apply the concept of complex neural networks do not provide the required amount of transparency. Medical workers will be skeptical about the accuracy of models and will not want to use automated prediction when the paths of decision are not understood [5]. The primary purpose of Explainable AI is to provide simple human-understandable explanations that authenticate model outputs and retain diagnostic accuracy [6], [7]. The XAI technology provides transparent explanations, which improve the credibility of AI systems to the staff and patients in healthcare [3], [8].

There are numerous challenges associated with these systems as researchers attempt to instantiate explainability features [4]. The healthcare sector is confronted with a persistent problem of defining the appropriate explanation due to the diverse groups of people such as clinicians and patients and regulators and developers requiring varying information volumes [5], [7]. Explainability has various approaches that range between post-hoc methods (SHAP values and saliency maps) to decision trees and rule-based systems that have an intrinsic interpretation [2], [6]. Researchers must determine the most trustworthy and useful as well as clinical value approaches. The clarity and correctness of the explanations is important as the complex information or the information that is confusing will not be trusted but rather diminished [1], [5].

## Literature Review

In the past few years, Explanable Artificial Intelligence (XAI) has become a crucial topic in research in the healthcare field. The primary benefit of XAI is not limited by the transparency of the algorithms since it allows medical personnel to trust technology systems and apply them to make improved treatment choices. The algorithms are very diagnostic but their lack of explainability makes them difficult to adopt by clinicians. The studies of XAI demonstrate promising outcomes and scientists have not overcome all its current difficulties.

The authors of Amann et al. [1] show the role of explainable systems in improving patient self-determination and holding doctors accountable. The study is unique in the sense that it provides the linkages among the legal frameworks and the ethical decision making systems. The study has theoretical orientation that does not allow the reader to know the practical approaches to medical work.

The authors of Brankovic et al. [2] prove that SHAP and LIME explanations do not substantively reduce the predictive modeling accuracy. Moreno-Sanchez [3] explores kidney disease detection diagnostic approaches by integrating explanatory tools into the diagnostic process. The study exhibits operational usefulness but with limited data sets that casts questions about its usefulness in the real hospital setting.

The studies have been devoted to the research of the impact of organizational structures on XAI adoption. The researchers of Raz et al. [4] explore the impact of healthcare organizations and the body of governing on XAI systems implementation. Rezaeian et al. [5] authors explore the impact of clear explanation on the level of medical staff trust. The study shows that the performance of the algorithms is not the only factor that defines the level of adoption since institutions and professional attitudes are also important. The study concentrates on the theory and particular cases as opposed to application.

There are research projects which engage in implementing XAI solutions beginning to emerge. AlpODS, a diagnostic system by Ultsch et al. [6] is a combination of accurate results and interpretable explanations. The Vani et al. [7] authors show examples of XAI use in the area of continuous health monitoring that allows providing patients with a personalized approach to medical care. The model developed by Wang et al. [8] is constructed on the principles of thinking and mimics the processes of clinical decision-making. The study is an expansion but an experimental one and in a small scale.

The current literature shows there has been constant growth but it remains challenged with several issues that are yet to be solved. The present study has three significant limitations since it is conducted on limited data and it is not validated by the tests or it does not examine the interaction of doctors and systems in real clinical settings. To attain a greater degree of reliability and practical value, the XAI of healthcare needs to be developed with the solutions to these current issues.

## Methods and Materials

This paper examined the attitudes of young practitioners in the medical field toward explainable artificial intelligence (XAI). A survey was conducted in the form of a structured online questionnaire, which included demographic items and questions on a Likert scale (1 = strongly disagree, 5 strongly agree) [9]. The survey involved several scales, including knowledge of XAI, confidence in AI-based decision-making, confidence in AI-aided diagnosis, perceived usefulness, comfort with transparency, and readiness to use XAI (Fig. 1). The survey items were also a reword of validated instruments in previous research on AI adoption in healthcare practices, which guaranteed content validity.

| Q\R | R1 | R2 | R3 | R4 | R5 | R |
|---|---|---|---|---|---|---|
| Q1 Gender | Male | Female | Female | Male | Female | Male |
| Q2 Age | 18–22 | 18–22 | 23–27 | 18–22 | 23–27 | 18–22 |
| Q3 Education Level | Pre-clinical | Pre-clinical | Clinical | Pre-clinical | Clinical | Pre-clinical |
| Q4 Field of study | General Med | Dentistry | Pharmacy | General Med | Dentistry | General Med |
| Q5 Experience with AI | Yes | No | Yes | No | Yes | Yes |
| Q6 I understand XAI | 5 | 4 | 5 | 4 | 5 | |
| Q7 Explanations help me understand | 4 | 5 | 4 | 4 | 5 | |
| Q8 I trust AI more with explanations | 5 | 4 | 5 | 5 | 4 | |
| Q9 Seeing AI decisions increases confidence | 4 | 3 | 5 | 4 | 5 | |
| Q10 XAI can reduce errors | 5 | 4 | 5 | 3 | 4 | |
| Q11 Comfortable using AI that explains | 4 | 4 | 5 | 2 | 4 | |
| Q12 XAI makes decisions transparent | 3 | 4 | 5 | 2 | 4 | |
| Q13 Trust AI more if reasoning understood | 4 | 5 | 3 | 4 | 5 | |
| Q14 Clearer explanation increases trust | 3 | 4 | 5 | 2 | 3 | |
| Q15 Transparency more important than speed | 4 | 3 | 5 | 2 | 4 | |
| Q16 Prefer doctor consultation | 3 | 2 | 5 | 2 | 3 | |
| Q17 Explanations make diagnosis safer | 5 | 2 | 4 | 5 | 4 | |
| Q18 Willingness to use XAI | 4 | 3 | 5 | 4 | 5 | |
| Q19 Training students is necessary | 5 | 3 | 4 | 5 | 4 | |

Fig. 1. Table with survey's responses

The convenience and snowball sampling of medical programs were used to recruit the participants [10]. The concluding sample contained 30 respondents who were aged between 18 and 27 years. The engagement was optional and informed consent was achieved by a checkbox during the first page of the online survey. To achieve qualitative triangulation, an interview with one professional who has experience with the use of AI in healthcare was conducted. In this mixed-method design, it was possible to integrate quantitative data of the survey with the contextual information of practice [11].

To determine the reliability of measurement scales, a pilot study was carried out. To determine internal consistency, the alpha of Cronbach was calculated, a standard reliability index for multi-item scales [12]. The alpha that was obtained was 0.702, which is sufficient to assume that the research was reliable enough to be used in the pilot study (Fig. 2). This proves that the items in the survey are a reliable measure of the desired constructs such as understanding, trust, confidence and perceived usefulness of XAI.

| | | | | | | |
|---|---|---|---|---|---|---|
| 22 | | 62 | 53 | 70 | 52 | 62 |
| 23 | | | | | | |
| 24 | | | | | | |
| 25 | | | | | | |
| 26 | Sum of All Variance --> | 31,68850575 | | Cronbach --> | | Sum of Variance --> |
| 27 | Each Variance --> | 0,6264367816 | | | | 10,55057471 |
| 28 | | 0,1885057471 | | 0,7021616575 | | |
| 29 | | 0,4643678161 | | | | |
| 30 | | 0,6022988506 | | | | |
| 31 | | 0,6540229885 | | | | |
| 32 | | 0,3264367816 | | | | |
| 33 | | 0,8517241379 | | | | |
| 34 | | 0,6896551724 | | | | |
| 35 | | 0,9609195402 | | | | |
| 36 | | 0,7540229885 | | | | |
| 37 | | 1,412643678 | | | | |
| 38 | | 1,385057471 | | | | |
| 39 | | 0,516091954 | | | | |
| 40 | | 0,5333333333 | | | | |
| 41 | | 0,5850574713 | | | | |

Fig. 2. Pilot Study Reliability Results (Cronbach's Alpha From Excel)

Table 1, Participants' Demographic Profile, gives the demographic makeup of the sample. The sample had an equal representation of boys and girls (half-male, half-female). Majority of the respondents were 18-22 (53%) and pre-clinical years (53%) old with others in clinical years (47%). Most of them were pursuing General Medicine (57%), with Dentistry, Pharmacy, and other small proportions (30, 13 respectively). Speaking of previous experience with AI, 57 percent of respondents said that they were familiar with AI tools, which can affect their perception and belief in AI-based decision-making.

Table 1. Participants' Demographic Profile

| Variable | Category | n | % |
|---|---|---|---|
| **Gender** | Male | 15 | 50% |
| | Female | 15 | 50% |
| **Age** | 18–22 | 16 | 53% |
| | 23–27 | 14 | 47% |
| **Education Level** | Pre-clinical (Years 1–3) | 16 | 53% |
| | Clinical (Years 4–6) | 14 | 47% |
| **Field of Study** | General Medicine | 17 | 57% |
| | Dentistry | 9 | 30% |
| | Pharmacy | 4 | 13% |
| **Experience with AI** | Yes | 17 | 57% |
| | No | 13 | 43% |

Descriptive and inferential statistics were used to analyze the data of the surveys. Participant responses to all survey questions were summarized using descriptive statistics (mean, median, standard deviation, minimum, maximum). Pearson correlation analysis was used in inferential statistics that analyzed the relationship among the key constructs in understanding, trust, confidence and willingness to adopt XAI [13]. This analysis made it possible to test the research hypothesis according to which better understanding and confidence with the AI systems is linked to better trust and adoption intentions. All the analysis was done through conventional statistical software.

**Results**

The survey data indicate that the attitudes towards Explainable AI (XAI) are mostly positive among medical students, with the mean scores on most items being above the middle of the scale (3.5) (Table 2). To illustrate, trust in AI in case of explanations is given got a mean of 4.47 (SD=0.72) whereas readiness to embrace XAI had a mean of 4.07 (SD=0.73). The standard deviations were low and moderate, which means that answers were quite similar and that participants had the same perceptions, to the greatest extent.

Table 2. Descriptive Statistics of Participants' Perceptions Toward Explainable AI

| Variable | Mean | Median | SD | Min | Max |
|---|---|---|---|---|---|
| Q6 | 4.17 | 4 | 0.74 | 2 | 5 |
| Q7 | 4.07 | 4 | 0.49 | 3 | 5 |
| Q8 | 4.47 | 5 | 0.72 | 3 | 5 |
| Q9 | 4.10 | 4 | 0.80 | 3 | 5 |
| Q10 | 4.07 | 4 | 0.81 | 3 | 5 |
| Q11 | 4.20 | 4 | 0.74 | 2 | 5 |
| Q12 | 4.00 | 4 | 0.92 | 2 | 5 |
| Q13 | 4.07 | 4 | 0.78 | 3 | 5 |
| Q14 | 4.13 | 4 | 0.81 | 2 | 5 |
| Q15 | 3.73 | 4 | 0.88 | 2 | 5 |
| Q16 | 3.60 | 3.5 | 1.14 | 2 | 5 |
| Q17 | 4.07 | 4 | 1.05 | 2 | 5 |
| Q18 | 4.07 | 4 | 0.73 | 3 | 5 |
| Q19 | 4.47 | 5 | 0.71 | 3 | 5 |
| Q20 | 4.07 | 4 | 0.87 | 3 | 5 |

The reliability of the measurement scales was established by carrying out a pilot study. The alpha of Cronbach of the constructs used to measure the understanding of XAI, the trust in AI-supported decisions, confidence in AI-assisted diagnoses, and the perceived usefulness were 0.702, which means that the alpha is acceptable in exploratory research. This ascertains that the scales are good in capturing the intended constructs and can be employed in supporting descriptive and inferential analyses.

A comparison of the results of the observed survey with the value that would have occurred according to the study hypotheses is given in Table 3. The majority of the items showed a great deal of alignment, especially those that were used to assess understanding, trust and perceived usefulness of XAI. As an example, the participants said that the explanations promoted a better understanding (Q7, mean = 4.07) and a higher level of confidence in the decisions made with the help of AI (Q8, mean = 4.47), which proves that explainability has a positive impact on the intention to adopt.

Table 3. Summary of Survey Responses, Expected Answers, and Alignment with Study Hypotheses

| Question | Real Answer (Most Frequent / Count) | Expected Answer | Follows Hypothesis (Yes/No) | Outcomes from Real Answers | Additional Feedback / Comments |
|---|---|---|---|---|---|
| Q1 Gender | Male & Female (15/15) | Female | Partially | Equal distribution of gender | Gender may affect perception of AI; consider in further analysis |
| Q2 Age | 18–22 (16/30) | 18–22 | Yes | Most respondents are younger students | Age influences understanding of XAI |
| Q3 Education Level | Pre-clinical (16/30) | Pre-clinical | Yes | Majority are at early stage of training | Less clinical experience may reduce trust in AI |
| Q4 Field of study | General Medicine (17/30) | General Medicin e | Yes | Most respondents are medical students | Field of study impacts XAI perception |
| Q5 Experience with AI | Yes (17/30) | Yes | Yes | Majority have prior AI experience | Experience increases understanding and trust in XAI |
| Q6 I understand XAI | 5 (14/30) | 4 | Yes | Students generally understand XAI concepts | Advanced training could further improve understanding |
| Q7 Explanations help me understand | 5 (24/30) | 4 | Yes | Explanations help most respondents | Interface clarity and explanation quality could be improved |
| Q8 I trust AI more with explanations | 5 (17/30) | 5 | Yes | Trust increases with explanations | Confirms that explainability is key for trust |
| Q9 Seeing AI decisions increases confidence | 5 (12/30) | 4 | Yes | Visual AI decisions boost confidence | Recommend using visual explanation tools |
| Q10 XAI can reduce errors | 5 (10/30) | 5 | Yes | Students believe XAI reduces errors | Confirms hypothesis about safety enhancement |
| Q11 Comfortable using AI that explains | 5 (23/30) | 4 | Yes | Most feel comfortable with explainable AI | Comfort depends on interface and prior experience |
| Q12 XAI makes decisions transparent | 5 (11/30) | 4 | Yes | Transparency of AI decisions is valued | Highlights the importance of explainability |
| Q13 Trust AI more if reasoning understood | 5 (20/30) | 5 | Yes | Understanding AI reasoning strongly increases trust | Key factor for adoption |
| Q14 Clearer explanation increases trust | 5 (17/30) | 5 | Yes | Clear explanations improve trust | Further interface improvement can enhance effect |
| Q15 Transparency more important than speed | 5 (18/30) | 4 | Yes | Students value transparency over speed | Matches expectations in educational context |
| Q16 Prefer doctor consultation | 5 (9/30) | 3 | No | Most still prefer doctor consultation | Hypothesis that AI could replace doctor is rejected |
| Q17 Explanations make diagnosis safer | 5 (13/30) | 5 | Yes | Explanations improve perceived safety | Supports practical value of XAI |
| Q18 Willingness to use XAI | 5 (15/30) | 4 | Yes | High willingness to use XAI | Positive potential for educational and clinical use |
| Q19 Training students is necessary | 5 (20/30) | 5 | Yes | Students consider training necessary | Supports development of educational programs for XAI |
| Q20 AI can improve doctor–patient interactions | 5 (13/30) | 4 | Yes | AI seen as a tool to enhance interactions | Confirms that AI can support doctors rather than replace them |

There was an exception of Q16 (preference to doctor consultation) in which most of the participants preferred human clinical advice to AI-only consultation. This observation suggests that although XAI is appreciated, it is mainly considered as a support system as opposed to medical judgment. The patterns below point to the subtlety of the role of XAI in healthcare, in which explainable systems are supposed to complement, but not substitute the expertise of clinical professionals.

Pearson correlation analysis was done to assess the relationships between major constructs (Table 4). The relationship between knowledge of XAI and belief in AI-aided choices had a moderate positive correlation ($r = 0.48$, $p = 0.007$), which means that more comprehensible the participants regarding AI reasoning, the higher were the chances to trust AI outputs. The confidence in AI-assisted diagnoses showed a positive significant correlation with perceived usefulness ($r = 0.60$, $p = 0.001$), which implies that higher the confidence people have in AI decisions, the more they perceive the usefulness of such decisions.

These results are empirical evidence that support the hypothesis that explainability is an important factor in user adoption. They show that understanding has a direct impact on trust that in effect shapes perceived utility and readiness to implement XAI into clinical processes. Additionally, the positive relationships suggest that an attempt to improve XAI interpretability would be compounding, and both trust and perceived usefulness would be improved at the same time.

The findings suggest that young medical workers tend to express positive sides of XAI. Descriptive statistics (Table 2) indicate that mean scores were above 3.7 on most constructs indicating that there was an agreement that XAI enhances the understanding, trust, confidence, and the perceived usefulness. The low standard deviations are also an indication of a common sense among the respondents, which increases the degree of trust in the generalizability of these pilot results on the population sampled.

These interpretations are corroborated by the results of correlation (Table 4). The correlation between knowledge and trust ($r = 0.48$) demonstrates that awareness of AI reasoning systems breeds trust in the use of AI in decision-making. The correlation between the two variables (confidence and perceived usefulness $r = 0.60$) demonstrates that the trust in AI implies the perception of practical utility, which implies the interdependence of cognitive understanding, emotional trust, and the desire to act.

Other clues are gained in answers about transparency and preference in consultation. The participants valued transparency more than speed (Q15) and did not lose their interest to human consultation (Q16) and these factors indicate a conceptualization of XAI as a decision support system instead of an autonomous system. This highlights an important point that XAI integration needs to be successful, with a focus on balancing between interpretability, usability, and clinician oversight to be consistent with professional expectations and ethical standards.

There are a number of limitations that should be considered. To begin with, the sample size ($N = 30$) is not that big, which restricts the statistical power and the possibility to apply the results to larger groups of individuals, especially those who have worked in various clinics or work in interdisciplinary teams. Second, the convenience and snowball sampling techniques open up the chance of possible selection bias; it is possible that the respondents are similar in terms of their educational background, familiarity with technology, or interest in AI which will overvalue positive perceptions.

Third, the researchers use self-reported responses only to the surveys. Although useful in perception research, such responses might not be representative of actual behavior since there is the social desirability bias or cognitive overestimation of understanding or hypothetical assessment of AI explanations. Fourth, the participants were not exposed to functional XAI systems. Consequently, findings are based on perceived and not practical usability and the actual problem in the environment, like interface constraints, complex explanations, time constraints, and varying environments might influence trust and adoption.

Fifth, the research analyzes a restricted number of constructs. Other powerful variables such as perceived risk, ethical issues, institutional policies, previous clinical training and actual patient interactions were not assessed. Sixth, the perceptions could be influenced by cultural and institutional setting; data were gathered within one medical educational institution and findings might be varied in other countries, healthcare systems, or cultures of professions.

**Discussion**

The findings prove that XAI is an essential element that has a positive impact on the perception of the users of the AI-based diagnostic tools [14].

Descriptive statistics (Table 2) reveal that the attitudes towards XAI are rather positive with mean scores exceeding 3.7 in the majority of constructs implying that respondents view XAI as increasing the level of understanding, trust, confidence, and perceived usefulness [15]. The standard deviations were rather low and moderate, which means that the perceptions were similar among the respondents. It is also interesting to note that the mean of trust in AI with explanations was 4.47 ($SD = 0.72$) and willingness to use XAI was 4.07 ($SD = 0.73$), which reflects the importance of explanations in clinical decision-making.

The consistency of the survey instruments was verified by conducting a pilot study where Cronbach's alpha ($a = 0.702$) was used showing that it has acceptable internal consistency to be used in exploratory research[16]. Constructs to determine XAI knowledge, trust in AI-informed decisions, confidence in AI-aided diagnoses, and perceived usefulness had adequate measurement properties, which allowed the validity of descriptive and inferential analysis to follow.

These relationships can be further explained by means of inferential analysis (Table 4). The outcomes of the Pearson correlation demonstrated that comprehension of XAI and confidence in the AI-assisted decisions had a significant positive relationship ($r = 0.48$, $p < 0.01$), meaning that the knowledge about AI rationality has a direct positive effect on trust in the results of the system [17]. Also, the perceived usefulness was linked with strong trust in AI-assisted diagnoses ($r = 0.60$, $p < 0.001$), which indicates that trust in AI can be converted to the perceived benefits of clinical decision-making in the clinical setting [18]. These results corroborate theoretical premises in the focus of cognitive comprehension and elucidation as precursors of technology adoption in healthcare.

Other insights are responses to questions about preferences on transparency and consultation. The participants cared more about transparency (Q15) than speed and retained their traditional attitude towards human consultation (Q16), indicating that XAI is still perceived rather as an auxiliary tool to help clinicians make decisions than as substituting them [19]. This highlights the significance of achieving a balance between interpretability, usability, and professional controls in terms of introducing XAI into medical procedures.

The research has practical implications in the healthcare organizations. Findings indicate that young medical professionals tend to embrace XAI systems more when explanations are easily comprehensible, understandable, and clinically meaningful. To make interpretability methods, including SHAP, saliency maps, and rule-based models, more comprehensible and trusted, training programs ought to be based on their interpretation [19], [20]. Furthermore, user-friendly interfaces must be considered by developers and hospitals because comprehensibility of the explanations is related to increased trust and confidence.

There are a number of limitations that should be taken into account. First, the sample size (N = 30) is small, which decreases the power of statistics and restricts the extrapolation of the findings, particularly to the experience of the clinician or an interdisciplinary team. Second, snowball and convenience sampling can result in selection bias because the respondents will probably have the same educational background, familiarity with technology, or interest in AI [20].

Third, only self-reported surveys were used in the study, which might not be applicable to real-life practice in clinical settings. Social desirability bias or overestimation of knowledge might result in responding to the questions. Fourth, the participants were not exposed to operational XAI systems, and therefore, the findings indicate perceived and not practical usability. Adoption and trust can be impacted by real-life issues (interface limitations, complexity of explanation, time, and contextual variability) [21].

Fifth, the research had a measurement of only a few variables. Other factors including perceived risk, ethical considerations, institutional policies, previous training and actual clinical experience were not evaluated but are likely to make a great impact on adoption behaviour. Lastly, results can be influenced by cultural and institutional context since information was gathered in one learning institution among young medical learners; cross cultural validation has to be carried out prior to making generalizations [22], [23].

To eliminate these shortcomings, it is recommended that future study should be increased in sample size and participant range including more clinicians in different career stages to help in generalizing the results. Stratified or randomized sampling techniques are proposed to reduce the possible selection bias and guarantee a more accurate depiction of the medical professionals. Because self-reported perceptions are not a good indicator of how people actually act, experimental designs, involving direct interaction with operational XAI systems, would be helpful in revealing the actual patterns of use and decision-making. Further research should also consider more of the measured variables to incorporate the ethical variables, institutional policies, previous clinical training, and other variables as they exist in practice to contribute to adoption. The longitudinal studies would also enable the researchers to follow the variations in perceptions and trust in case the participants are able to have real-life experience with XAI, and this would provide a more detailed insight into how the process of adopting changes across time.

Irrespective of these shortcomings, the research offers initial empirical findings that justify explainability, which are increasing trust, confidence, and the desire to use AI in the medical setting. It provides practical knowledge on how to create XAI systems and interventions in the form of educational initiatives, which will facilitate the safe and efficient application of explainable AI in clinical decision-making.

## Conclusion

The study examined the effectiveness of Explainable Artificial Intelligence (XAI) systems on trust and transparency in medical diagnosis systems through the lenses of the opinions of young medical professionals. The results prove that explainability is a necessary condition to the successful implementation of AI in healthcare. As the study shows, explanation of AI predictions results in increased trust, confidence, and more acceptance of AI-assisted diagnoses.

The visual, textual, or feature-based explanation allows users to gain more insight into the model decisions, become more capable of identifying the possible mistakes, and have more confidence in using AI tools. This is in line with existing literature that shows that uninterpretable black-box models can be readily rejected in clinical practice since they lack transparency and may contain hidden biases. By providing clear explanations, XAI technology will also lower the level of uncertainty and allow medical practitioners to make more informed decisions.

Statistical modeling application indicated that trust is the most important determinant of adoption, with quality and understanding of the explanations being the determinants. Thus, user-centered explainability needs to be on the agenda of medical AI systems, and its explanations should not only be accurate but also meaningful and relevant to clinical decision-making.

Although very informative, this research is limited in many ways, the first one is a small sample size, the homogeneous population of the participants, and the use of survey responses to obtain the information about XAI instead of direct interaction with the functional XAI tools. Further studies should consider the XAI functionality using experimental research with real-world application of AI tools with different clinical groups and clinicians with experience. The perceived risk, the quality of training, the readiness of the institution, and ethical implications are other aspects of AI adoption that should be analyzed to gain a deeper insight into patterns of AI adoption.

Altogether, this paper gives empirical findings that artificial intelligence systems need clarifying mechanisms to guarantee safe and ethical medical use. XAI is an essential prerequisite of AI-aided medical decision-making by establishing a trust basis, guaranteeing transparency, and improving diagnostic accuracy. Clinical AI deployment is achieved through explainability features that are deployed in an integrated manner; this feature is crucial in maximizing the advantages of this technology to involved stakeholders (the healthcare giver and the patient).

## Appendices / Supplementary Materials

Appendix A – Full Survey Data
The full dataset of survey responses, including demographic information, Likert-scale answers, and scores for all measured constructs. Accessible online via Google Sheets:
https://docs.google.com/spreadsheets/d/1ZpTs6YkAociFgWarnBvCbK7hJQty5xoUMnib7xyWz0o/edit?usp=sharing

## Implementation of the Questionnaire (with statistical results)

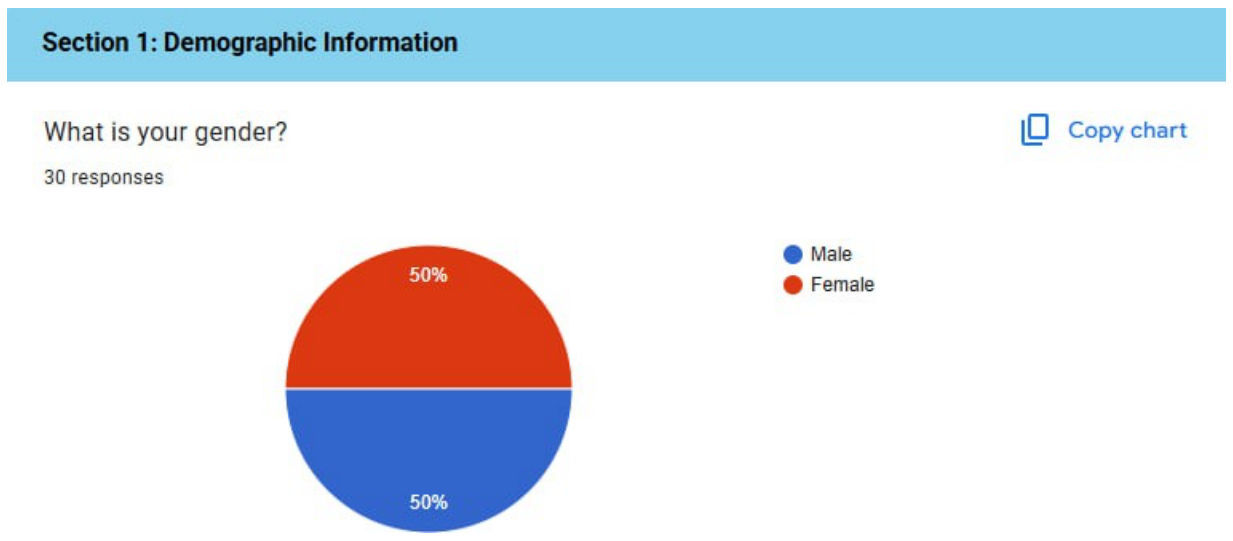

Fig. 3. The 1st question

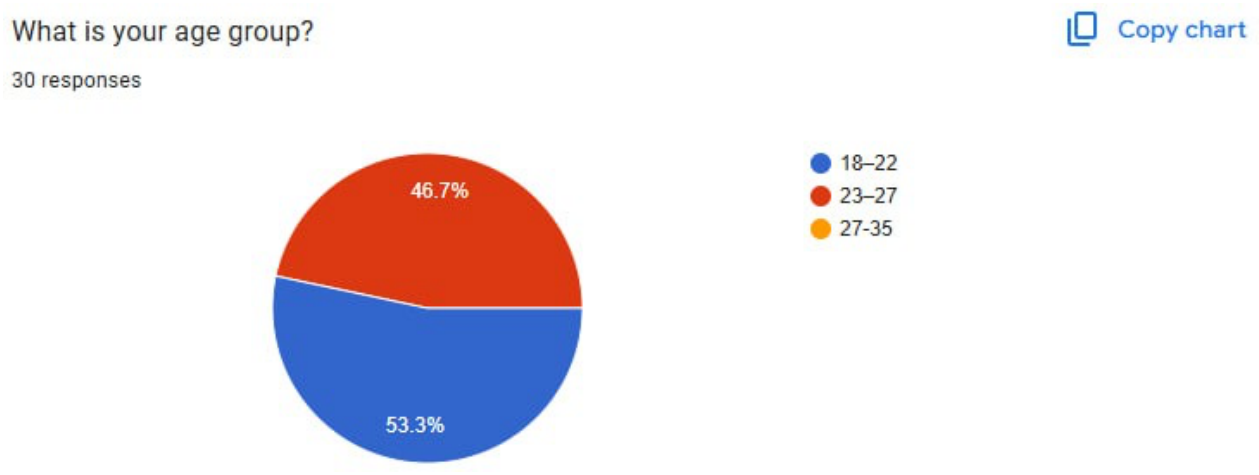

Fig. 4. The 2nd question

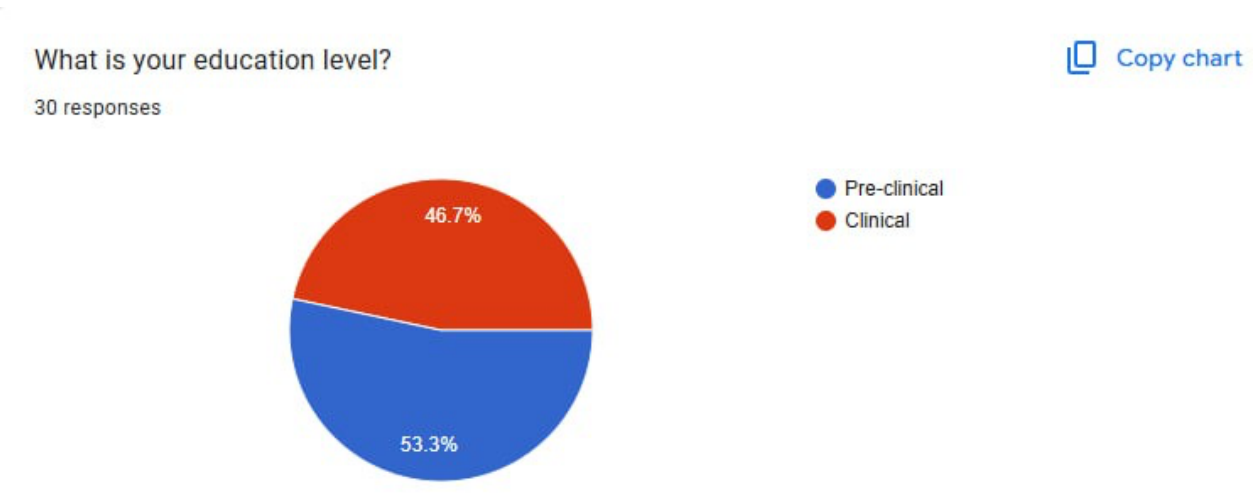

Fig. 5. The 3rd question

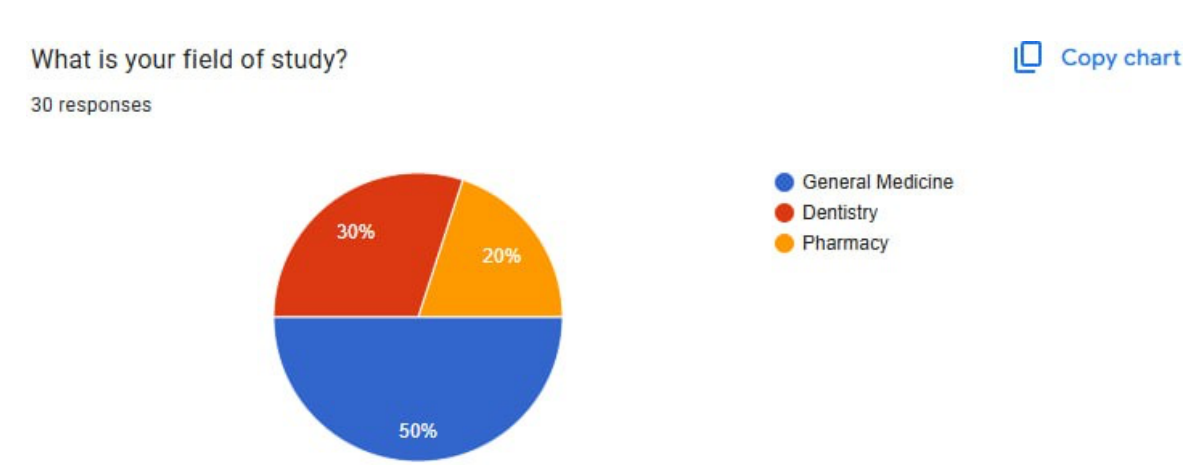

Fig. 6. The 4th question

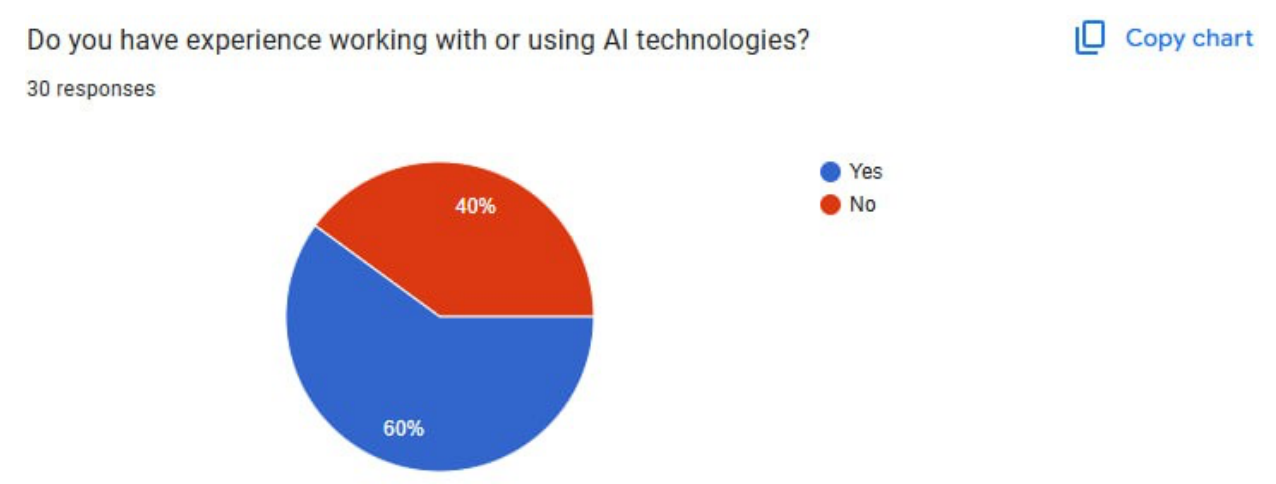

Fig. 7. The 5th question

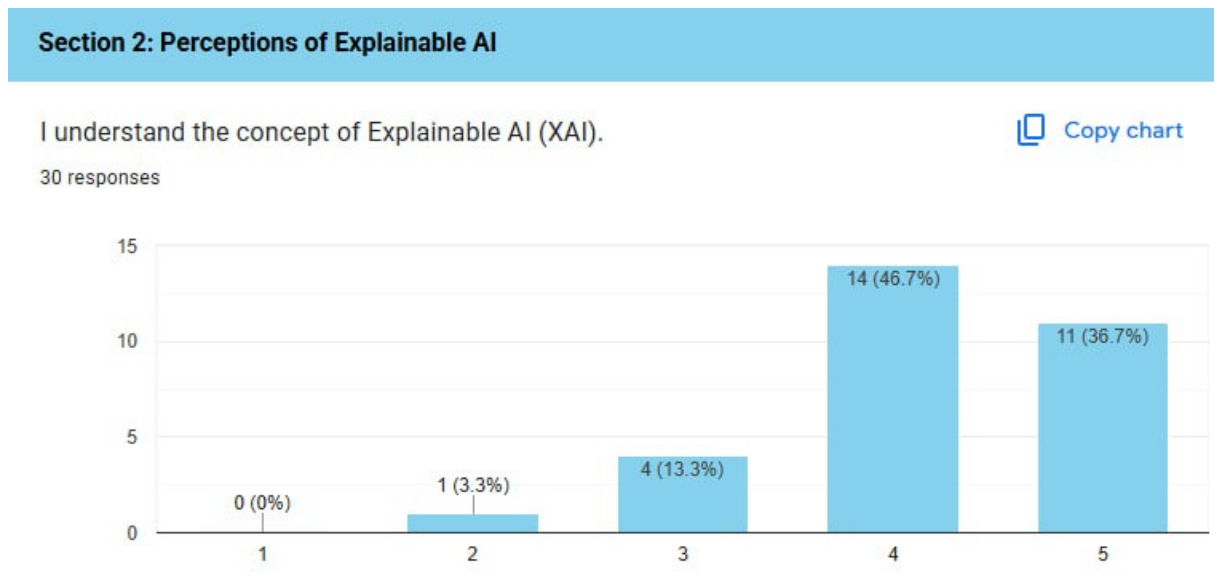

Fig. 8. The 6th question

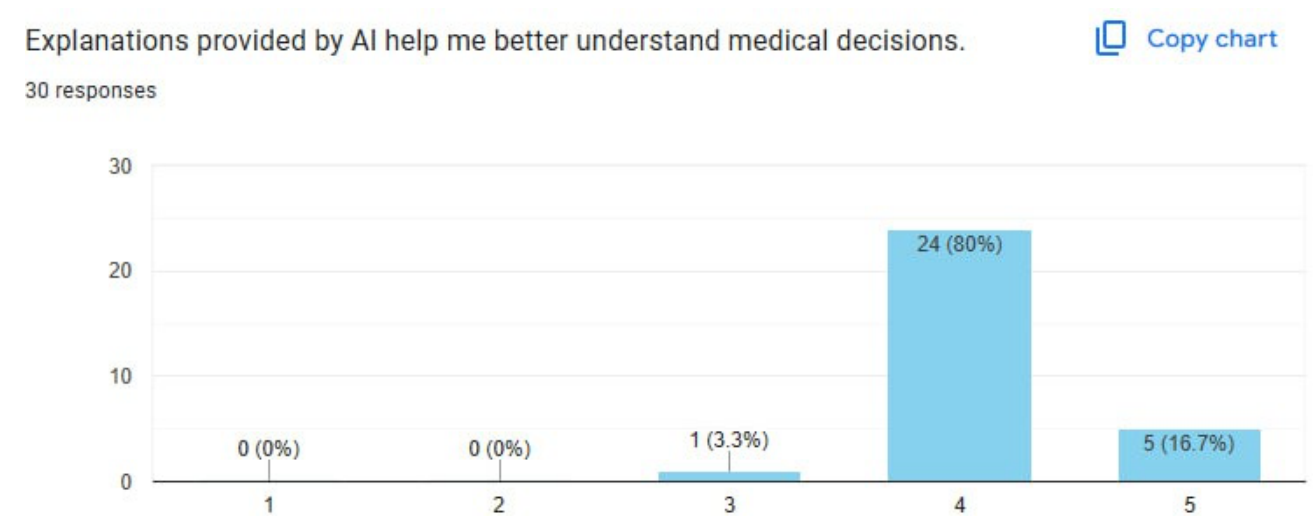

Fig. 9. The 7th question

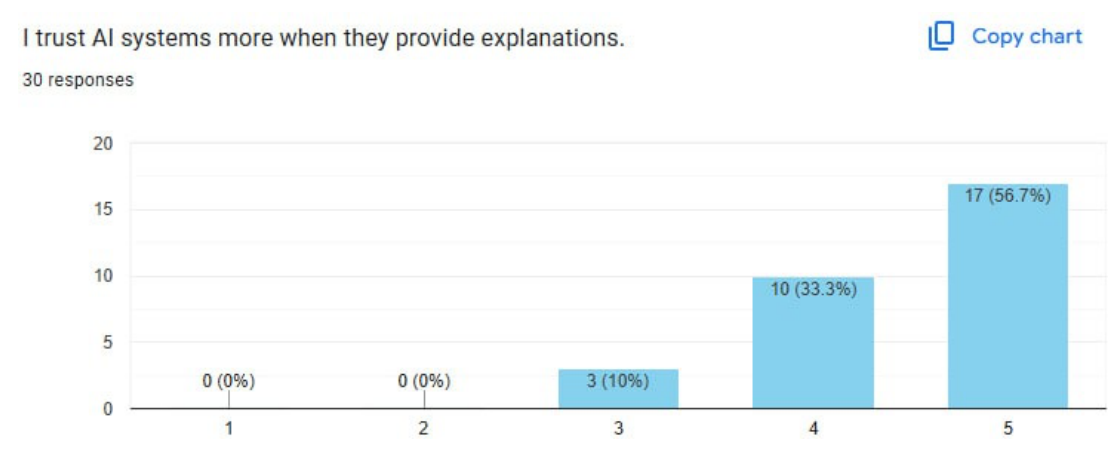

Fig. 10. The 8th question

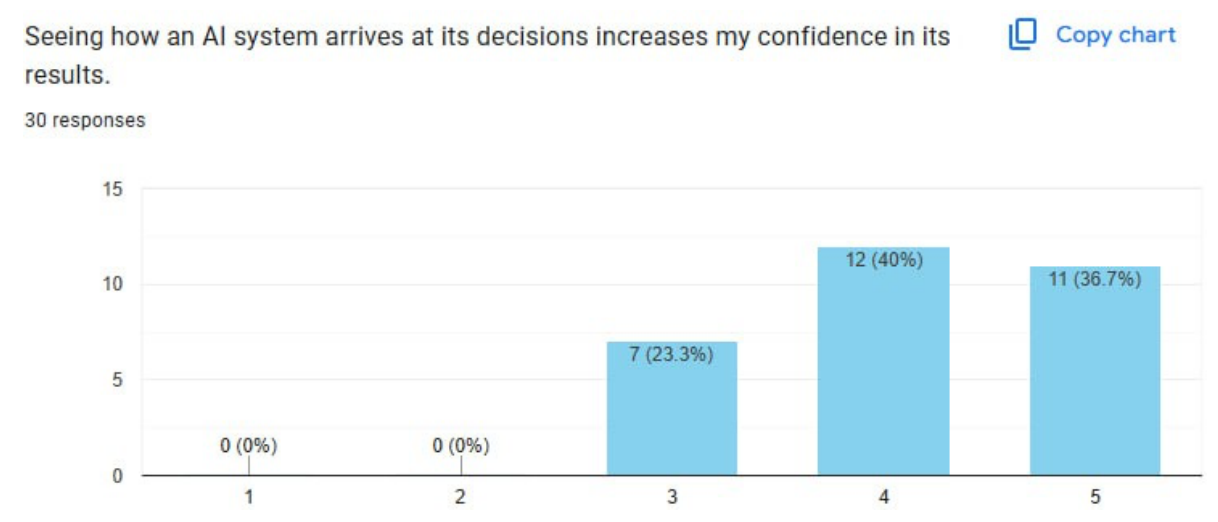

Fig. 11. The 9th question

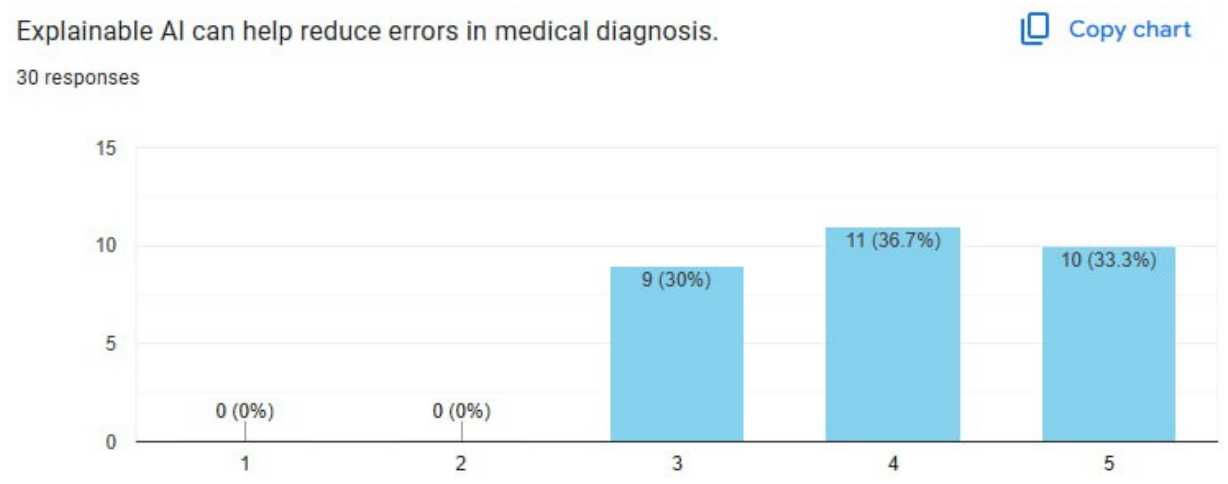

Fig. 12. The 10th question

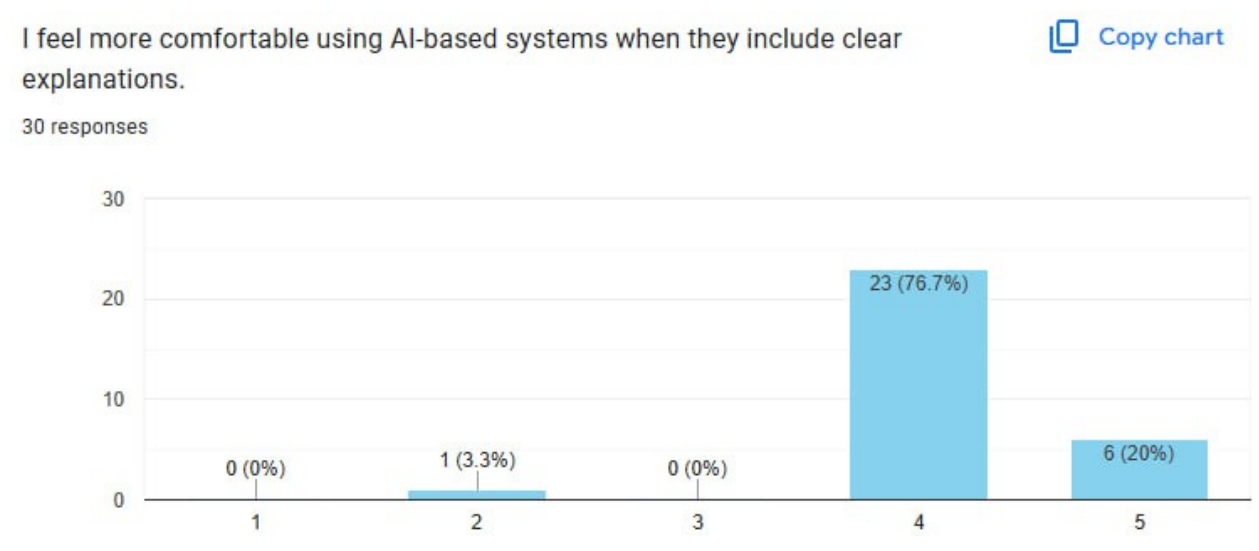

Fig. 13. The 11th question

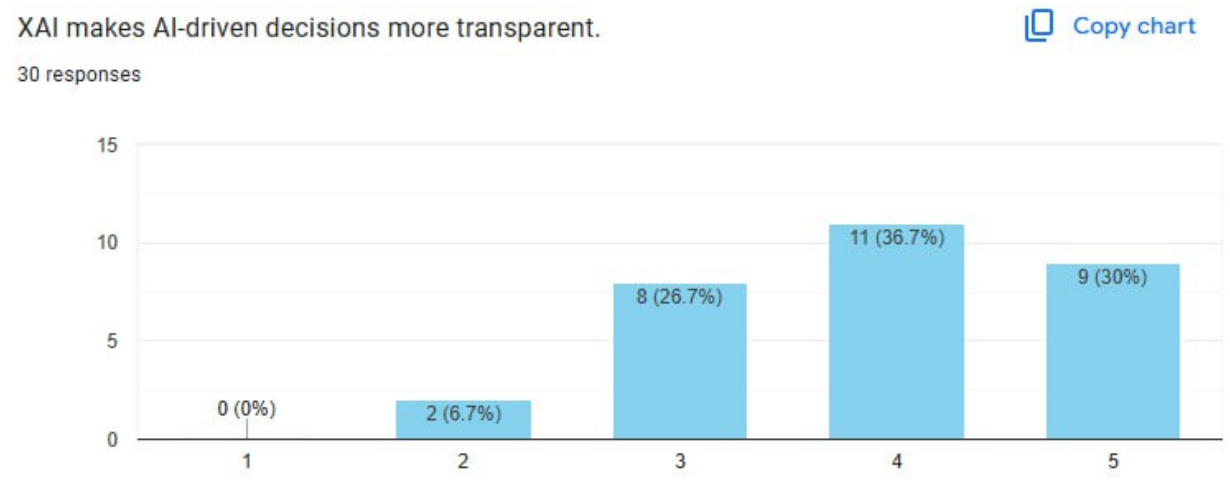

Fig. 14. The 12th question

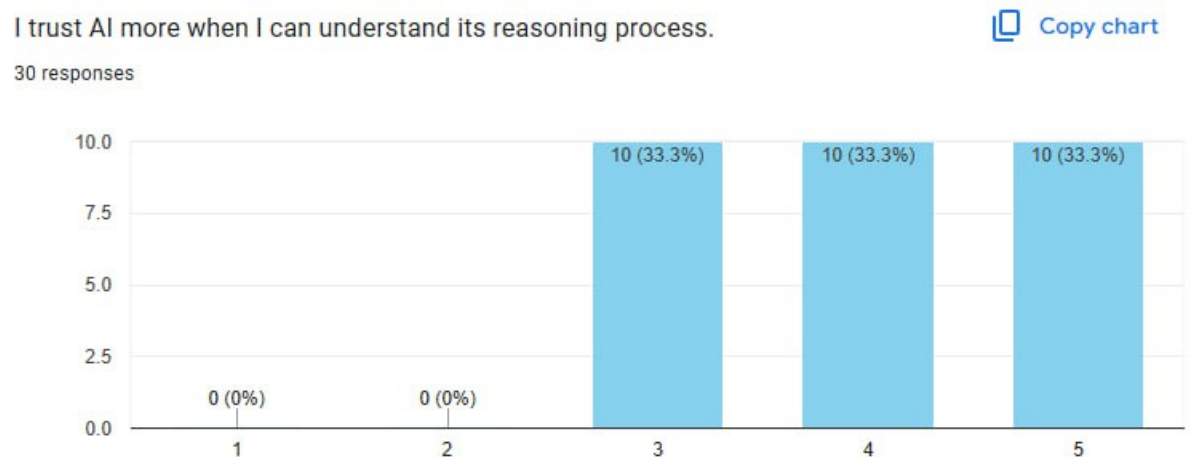

Fig. 15. The 13th question

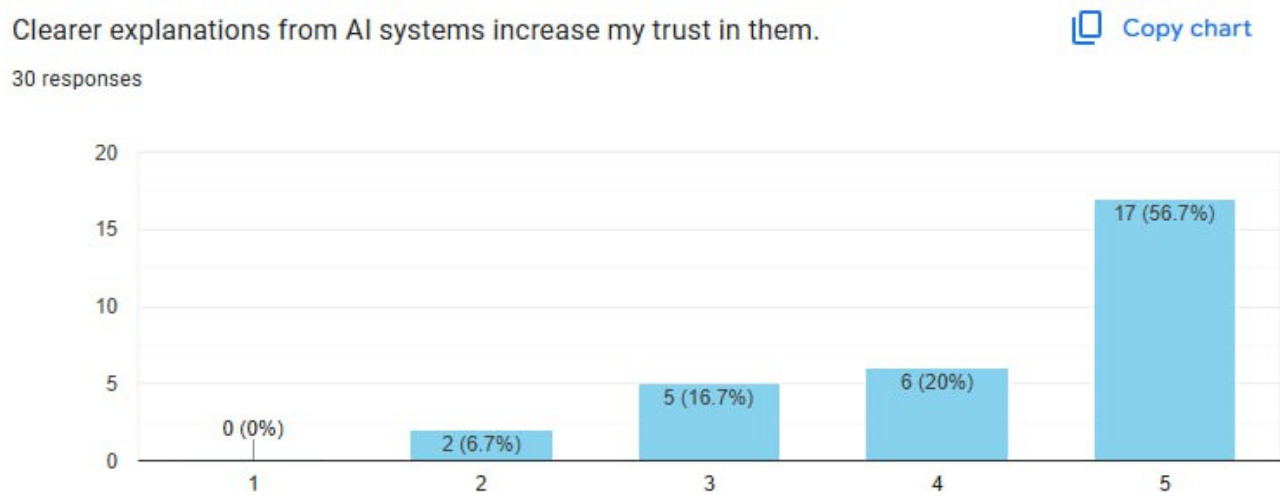

Fig. 16. The 14th question

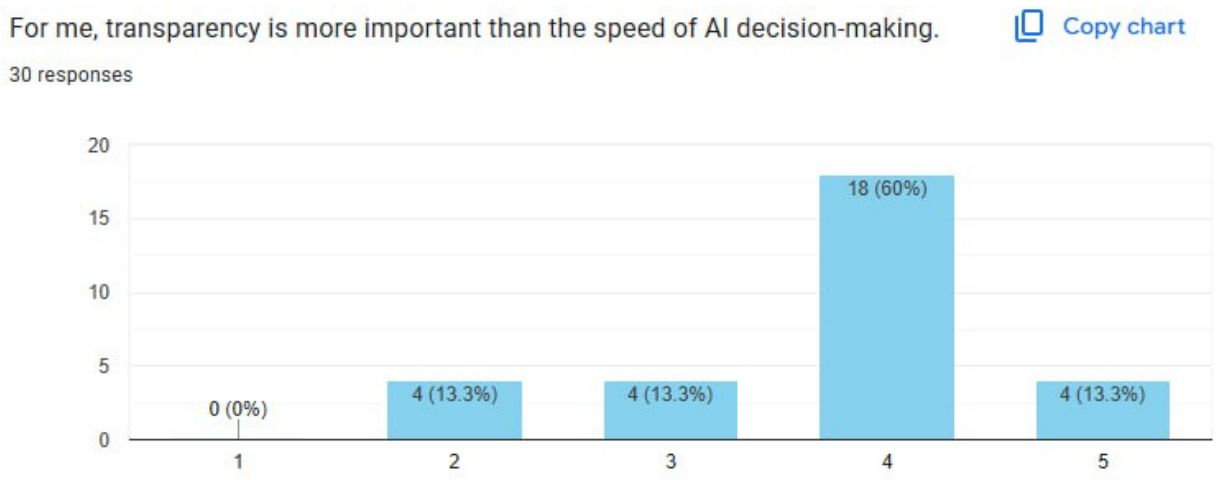

Fig. 17. The 15th question

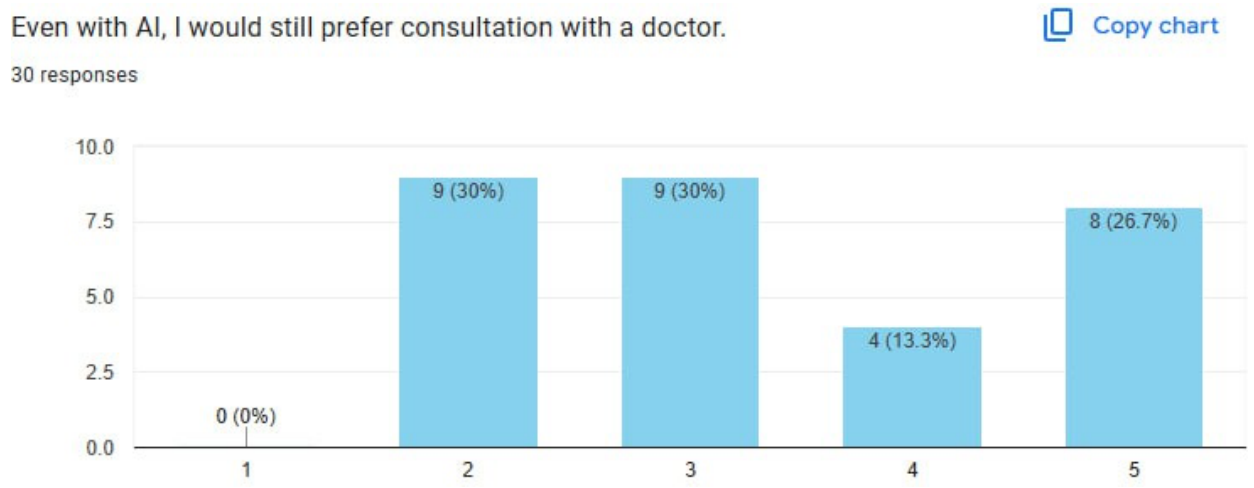

Fig. 18. The 16th question

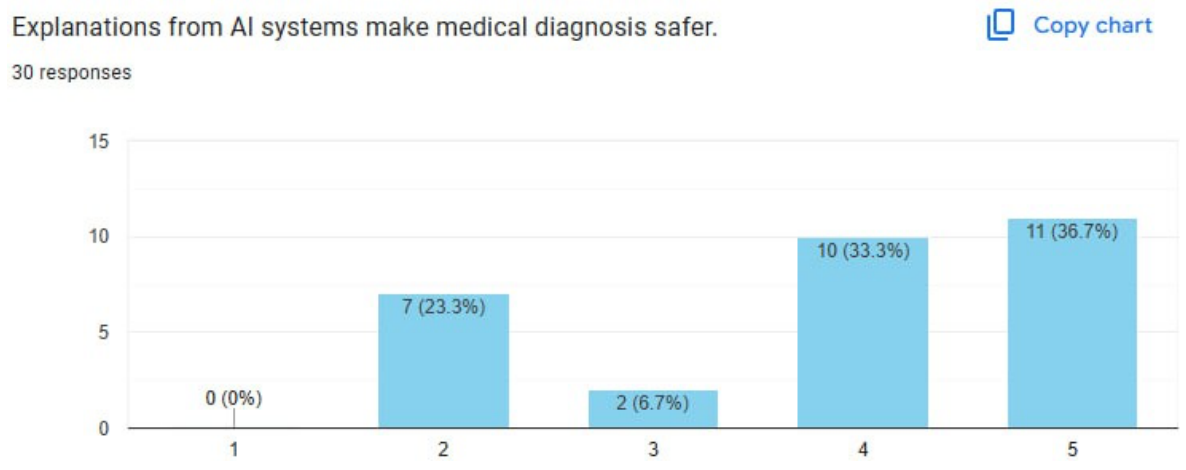

Fig. 19. The 17th question

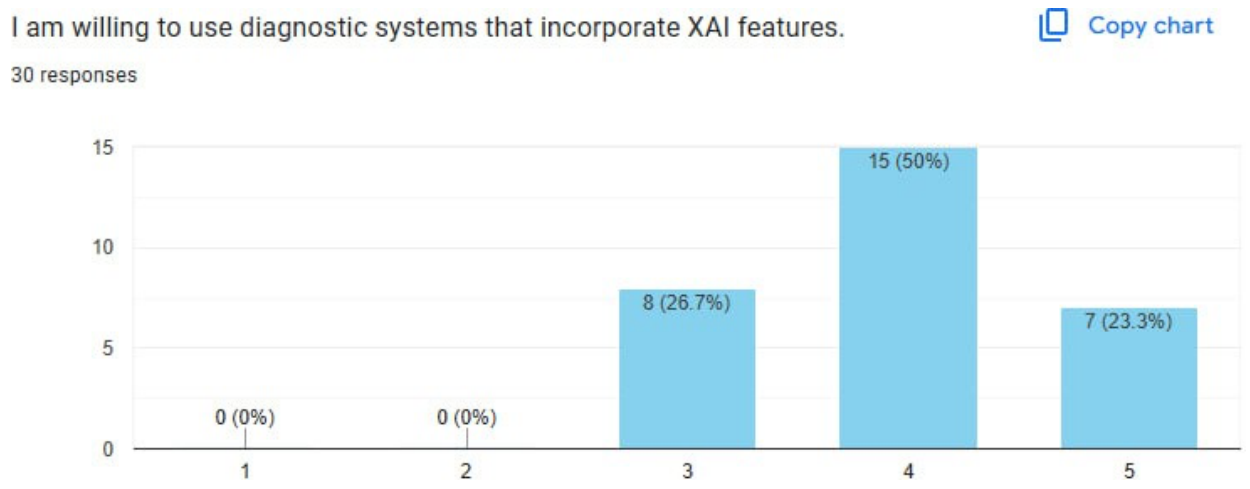

Fig. 20. The 18th question

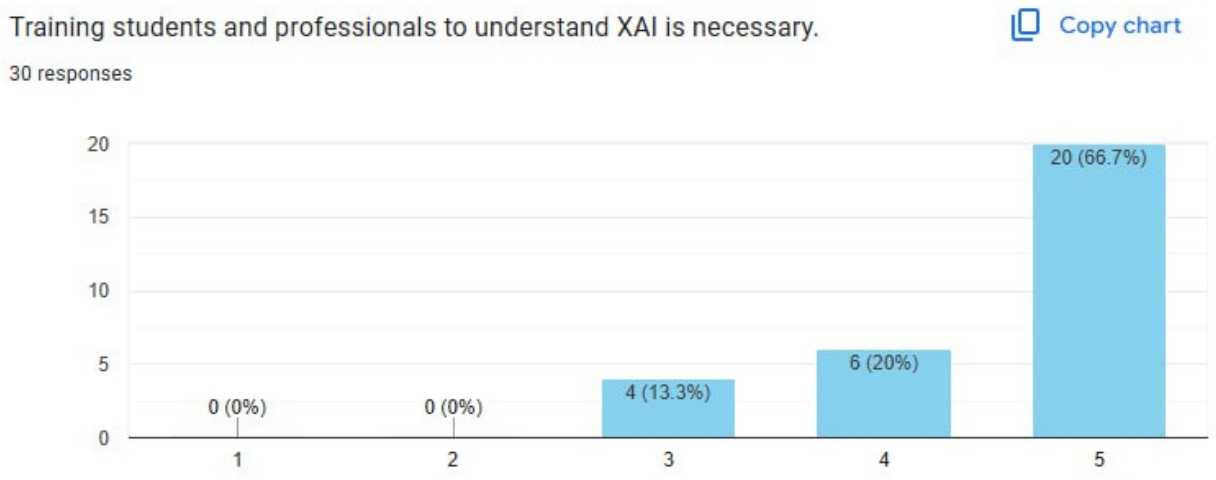

Fig. 21. The 19th question

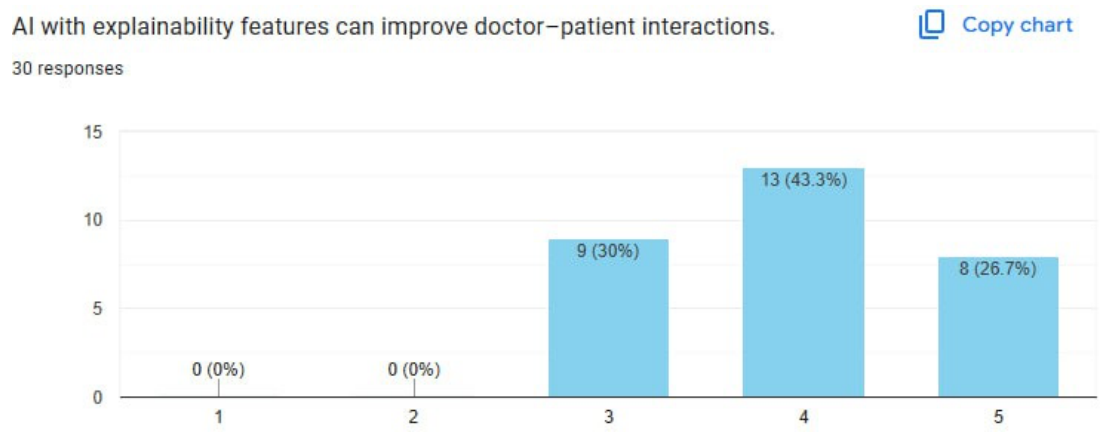

Fig. 22. The 20th question